\documentclass[aps,prd,preprint,amsmath,eqsecnum,citeautoscript,longbibliography]{revtex4-1}
\usepackage[utf8]{inputenc}
\newcommand{\bea}{\begin{eqnarray}}
\newcommand{\eea}{\end{eqnarray}}
\newcommand{\be}{\begin{equation}}
\newcommand{\ee}{\end{equation}}
\begin{document}
\title{Analytical study of a Kerr-Sen black hole and a charged massive scalar field}
\author{Canisius Bernard}
\email{canisius.bernard@gmail.com}
\affiliation{Center for Theoretical Physics, Department of Physics, Parahyangan Catholic University, Bandung 40141, Indonesia}
\date{\today}
\begin{abstract}
It is reported that Kerr-Newman and Kerr-Sen black holes are unstable to perturbations of charged massive scalar field. In this paper, we study analytically the complex frequencies which characterize charged massive scalar fields in a near-extremal Kerr-Sen black hole. For near-extremal Kerr-Sen black holes and for charged massive scalar fields in the eikonal large-mass $\mathcal{M}\gg \mu$ regime, where $\mathcal{M}$ is the mass of the black hole, and $\mu$ is the mass of the charged scalar field, we have obtained a simple expression for the dimensionless ratio $\omega_I /\left(\omega_R - \omega_c\right)$, where $\omega_I$ and $\omega_R$ are, respectively, the imaginary and real parts of the frequency of the modes, and $\omega_c$ is the critical frequency for the onset of super-radiance. We have also found our expression is consistent  with the result of Hod \cite{Hod:2016bas} for the case of a near-extremal Kerr-Newman black hole and the result of Zouros and Eardly \cite{Zouros:1979iw} for the case of neutral scalar fields in the background of a near-extremal Kerr black hole.
\end{abstract}
\maketitle
\section{Introduction}
\label{sec:intro}
Black holes are very exotic objects that are believed to exist in the Universe and should be studied both theoretically and experimentally. Despite their exotic nature, they obey the fundamental laws of physics, for instance, the law of conservation of energy and the second law of thermodynamics, which states that the total entropy of the Universe is always increasing as time passes.\\
\indent In Einstein's theory of gravity, black holes are often classically thought of as dead. They can absorb radiation and energy but cannot emit it \cite{Hawking:1999dp}. Moreover, naively, they do not seem to have an entropy. However, this leads to a paradox which was resolved by Bekenstein \cite{Bekenstein:1973ur} and further clarified by Hawking \cite{Hawking:1976de}. In particular, in 1969, Penrose \cite{Penrose:1969pc} contrived a classical process to extract energy from a rotating black hole. Another way to extract energy from a black hole is through the super-radiant effect and the instability of particles scattered from the black hole.\\
\indent The physical system we consider consists of a charged massive scalar field coupled to a charged rotating Kerr-Sen black hole. Attractively, this stationary scalar field configuration marks the physical boundary between a stable and unstable composed black-hole-charged massive-scalar-field stationary configuration. The black-hole-charged massive-scalar-field stationary configuration exist due to two distinct physical effects which together combine to hold the fields in the spacetime region exterior to the black hole horizon \cite{Hod:2014baa}. The first physical mechanism that is liable for the existence of these composed black-hole-charged massive-scalar-field stationary configurations is the phenomenon of super-radiant scattering of the scalar fields in the black hole background. In the case of rotating and charged black holes, it was discovered that scalar waves coming from infinity were attracted by the gravity of the black hole and scattered back to infinity with a scalar wave reflection coefficient greater than one if the scalar-field frequency is within the range 
\be \label{IN01}
0<\omega<m\Omega_H +q\Phi_H,
\ee
where $m$ is the momentum angular of the scalar field, $q$ is the charge of the scalar field, ${\Omega _H}$ is the angular velocity, and ${\Phi _H}$ is the electrostatic potential defined in Eqs. (\ref{KSG15}) and (\ref{KSG16}).
The second physical mechanism needed for the existence of these composed  black-hole-charged massive-scalar-field stationary configurations is provided by the mutual gravitational attraction between the black hole and the massive scalar field. A mass $\mu$ for a scalar field will forbid waves of frequency,	\be \label{IN02}
\omega^2 < \mu^2,
\ee
from escaping to infinity. Rather they will be reflected back into the black hole by a potential barrier. If the scalar field satisfies the condition (\ref{IN01}), i.e., super-radiance condition, the scalar wave will be amplified by the rotational energy of the black hole.\\
\indent Recent analytical and numerical studies \cite{Benone:2014ssa,Bernard:2016wqo,Delgado:2016jxq,Delgado:2016jxq,Degollado:2013eqa,Herdeiro:2014goa,Herdeiro:2015waa,Herdeiro:2015gia,Herdeiro:2014ima,Hod:2012px,Hod:2013eea,Hod:2014baa,Hod:2017vld,Huang:2016qnk,Huang:2017whw,Hwang:2011mn,Hansen:2015dxa,Hansen:2013vha,Hansen:2014rua,Jusufi:2016hcn,Kokkotas:2015uma,Konoplya:2011qq,Li:2015bfa,Li:2013jna,Nakonieczna:2015umf,Nakonieczna:2016iof,Sakalli:2016xoa,Sakalli:2016fif,Sakalli:2016jkf,Sakalli:2015uka,Siahaan:2015xna,Wu:2003qb} of the scalar fields coupled to black holes have revealed that charged (rotating) black holes can support stationary scalar configurations, which are regular at the horizon and outside it. These stationary clouds are the linear version of nonlinear hair \cite{Herdeiro:2014goa,Herdeiro:2015waa,Herdeiro:2015gia,Herdeiro:2014ima}.\\
\indent In a recent paper, we investigated stationary scalar clouds around a Kerr-Sen black hole \cite{Bernard:2016wqo}. By using a numerical method, Huang, Liu, Zhai, and Li have shown a numerical analysis of charged massive scalar clouds around Kerr-Sen black holes beyond the near-extremal limit \cite{Huang:2017whw}. The existence of the charged massive scalar fields in a Kerr-Sen black hole background should have an asymptotically exponentially decaying radial solution behavior \cite{Benone:2014ssa}. We found that this asymptotic radial solution matched this requirement well:
\be \label{IN03}
R_{lm}\left( r \right) \approx \left\{ \begin{array}{l}
	{e^{ - i\left( {\omega  - {\omega _c}} \right){r_*}}} ~~~~~{\rm{ for }}~r \to {r_ + },\\
	{\frac{1}{r}e^{ -{r}\sqrt {{\omega ^2} - {\mu ^2}} }} ~~~{\rm{ for }}~r \to \infty ,{\rm{ }}
\end{array} \right.
\ee 
where $\omega_c$ is the critical frequency
\be \label{IN04}
{\omega _c} \equiv m{\Omega _H} + q{\Phi _H}.
\ee
Here, $r_+$ is the radial coordinate of the outer horizon defined in Eq. (\ref{KSG08}). The ``tortoise'' radial coordinate $r_*$ is defined by
\be \label{IN05}
\frac{{d{r_*}}}{{dr}} = \frac{r\left(r+2b\right)+a^2}{{{\Delta _{KS}}}}.
\ee
The rotational parameter $a$ is defined as a ratio between the black hole's angular momentum $J$ to its mass $\cal{M}$. Both $b$ and $\Delta_{KS}$ are defined in Eqs. (\ref{KSG05}) and (\ref{KSG06}).\\
\indent The super-radiant instability of charged massive scalar fields in Kerr-Newman spacetime was studied in the important work by Furuhashi and Nambu in \cite{Furuhashi:2004jk}. It was found that, in the region satisfying the small frequency $\mathcal{M}\mu\ll 1$ and the small charged $|qQ|\ll 1$ regime, the imaginary parts of the frequencies which characterized charged massive scalar fields in the Kerr-Newman background are given by
\be \label{IN06}
\Im\omega  \approx \frac{{{\mu ^4}}}{{24}}{\left| {\mathcal{M}\mu  - qQ} \right|^5}a\left( {{\mathcal{M}^2} - {Q^2}} \right).
\ee
The purpose of this paper is analytically study the Kerr-Sen black hole and a charged massive scalar field.\\
\indent The paper is organized as follows. In Sec. \ref{sec:KSG}, we review the Kerr-Sen geometry and the corresponding charged massive-scalar-field equation. In Sec. \ref{sec:NFOreg}, we study and derive the radial solution in the far, near, and overlap regions. In Sec. \ref{sec:superradiant}, we derive the frequencies of the charged massive scalar fields in the background of a near-extremal Kerr-Sen black hole, and then in Sec. \ref{sec:eikonal}, we compare our large-mass regime result for the frequencies of the composed black-hole-field system with the corresponding result of Hod in \cite{Hod:2016bas} and Zouros and Eardly in \cite{Zouros:1979iw}. In the last section, we summarize the paper. In this paper, we use the units where $G=\hbar=c=1$.
\section{Charged Massive Scalar Field in Kerr-Sen Spacetime}
\label{sec:KSG}
In 1992, Sen obtained a four-dimensional charged and rotating black hole solution in the low-energy limit of heterotic string theory. The effective action in four dimensions reads  \citep{Sen:1992ua}
\be \label{KSG01}
S = \int {{d^4}x\sqrt {|\tilde{g}|} } {e^{ -  \Phi }}\left( {R - \frac{1}{8}{F_{\mu \nu }}{F^{\mu \nu }} + {\tilde{g}^{\mu \nu }}{\partial _\mu } \Phi {\partial _\nu } \Phi  - \frac{1}{{12}}{H_{\kappa \lambda \mu }}{H^{\kappa \lambda \mu }}} \right),
\ee
where $\tilde{g}$ is the determinant of the tensor metric $\tilde{g}_{\mu\nu}$\footnote{The spacetime metric in the Einstein frame $g_{\mu\nu}$ and string one $\tilde{g}_{\mu \nu}$ are related by $g_{\mu\nu}=e^{ - \Phi }\tilde{g}_{\mu \nu }$.}, $R$ is the scalar curvature, $ \Phi$ is the dilaton field, and $F_{\mu\nu}$ is the field-strength tensor
\be \label{KSG02}
{F_{\mu \nu }} = {\partial _\mu }{A_\nu } - {\partial _\nu }{A_\mu },
\ee
with $A_{\nu}$ being the electromagnetic 4-vector potential of the charged black hole. ${H_{\kappa \lambda \mu }}$ is the third-rank tensor field
\be \label{KSG03}
{H_{\kappa \mu \nu }} = {\partial _\kappa }{B_{\mu \nu }} + {\partial _\nu }{B_{\kappa \mu }} + {\partial _\mu }{B_{\nu \kappa }} - \frac{1}{4}\left( {{A_\kappa }{F_{\mu \nu }} + {A_\nu }{F_{\kappa \mu }} + {A_\mu }{F_{\nu \kappa }}} \right),
\ee
and $B_{\nu\sigma}$ is a second-rank antisymmetric tensor gauge field. Sen applied a transformation to the Kerr solution, known as a solution to the vacuum Einstein equation, to obtain the charged rotating black hole solution in the theory (\ref{KSG01}), known as the Kerr-Sen solution. In Boyer-Lindquist coordinates ($t,r,\theta,\phi$), the Kerr-Sen metric in the Einstein frame can be read as \cite{Ghezelbash:2012qn,Siahaan:2015xna,Siahaan:2015ljs,Bernard:2016wqo}
\bea \nonumber
d{s^2} &=&  - \left( {1 - \frac{{2\mathcal{M}r}}{{{\Sigma}}}} \right)d{t^2} + {\Sigma}\left( {\frac{{d{r^2}}}{{{\Delta _{KS}}}} + d{\theta ^2}} \right) - \frac{{4\mathcal{M}ra}}{{{\Sigma}}}{\sin ^2}\theta dtd\phi  \\\label{KSG04}&\quad&+ \left( {{\Sigma} + {a^2}{{\sin }^2}\theta  + \frac{{2\mathcal{M}r{a^2}{{\sin }^2}\theta }}{{{\Sigma}}}} \right){\sin ^2}\theta d{\phi ^2},
\eea
where
\be\label{KSG05}
b = \frac{{{Q^2}}}{{2\mathcal{M}}},
\ee
\be \label{KSG06}
\Delta _{KS}  = r({r} + 2b) - 2\mathcal{M}r + {a^2},
\ee
\be \label{KSG07}
\Sigma  = r({r} + 2b) + {a^2}{\cos ^2}\theta .
\ee
In fact, $r_+$ and $r_-$ represent the outer and inner black hole horizons given by  
\be \label{KSG08}
{r_ \pm } = \mathcal{M} - b \pm \sqrt {{{\left( {\mathcal{M} - b} \right)}^2} - {a^2}}.
\ee
The nonvanishing components of the Kerr-Sen contravariant tensor metric in the Einstein frame are
\be \nonumber
{g^{tt}} = \frac{{{\Delta _{KS}}{a^2}{{\sin }^2}\theta  - {{\left( {{r^2}+2br + {a^2}} \right)}^2}}}{{{\Delta _{KS}}\Sigma }},~{g^{rr}} = \frac{{{\Delta _{KS}}}}{\Sigma },
\ee
\be \nonumber
{g^{\theta \theta }} = \frac{1}{\Sigma },~{g^{\phi \phi }} = \frac{{{\Delta _{KS}} - {a^2}{{\sin }^2}\theta }}{{{\Delta _{KS}}\Sigma {{\sin }^2}\theta }},
\ee
\be\label{KSG09}
{g^{t\phi }} = {g^{\phi t}} =  - \frac{2\mathcal{M}ar}{{{\Delta _{KS}}\Sigma }},
\ee
with $\sqrt{-g}=\Sigma \sin \theta$. The Kerr-Sen metric (\ref{KSG04}) describes a black hole with mass $\mathcal{M}$, electric charge $Q$, and angular momentum $J=\mathcal{M}a$. The solutions for nongravitational fundamental fields in the theory described by the action (\ref{KSG01}) are 
\be \label{KSG10}
e^{-2 \Phi}  =  \frac{{{\Sigma}}}{{{r^2} + {a^2}{{\cos }^2}\theta }},
\ee
\be\label{KSG11}
{A_t} =  - \frac{{Qr}}{{{\Sigma}}},
\ee
\be \label{KSG12}
{A_\phi } = \frac{{Qra{{\sin }^2}\theta }}{{{\Sigma}}},
\ee 
\be \label{KSG13}
{B_{t\phi }} = \frac{{bra{{\sin }^2}\theta }}{{{\Sigma}}},
\ee 
and the related Hawking temperature, angular velocity, and electrostatic potential at the horizon are given by
\be \label{KSG14}
{T_H} = \frac{{{r_ + } - {r_ - }}}{{8\pi \mathcal{M}{r_ + }}}=\frac{{\sqrt {{{\left( {2{\mathcal{M}^2} - {Q^2}} \right)}^2} - 4{J^2}} }}{{4\pi \mathcal{M}\left( {2{\mathcal{M}^2} - {Q^2} + \sqrt {{{\left( {2{\mathcal{M}^2} - {Q^2}} \right)}^2} - 4{J^2}} } \right)}},
\ee 
\be \label{KSG15}
{\Omega _H} = \frac{a}{{2\mathcal{M}{r_ + }}}=\frac{J}{{\mathcal{M}\left( {2{\mathcal{M}^2} - {Q^2} + \sqrt {{{\left( {2{\mathcal{M}^2} - {Q^2}} \right)}^2} - 4{J^2}} } \right)}},
\ee
\be \label{KSG16}
{\Phi _H} = \frac{Q}{{2\mathcal{M}}}.
\ee
\indent Setting $b=0$, solutions for nongravitational fields (\ref{KSG10})--(\ref{KSG13}) vanish, and therefore, the line element (\ref{KSG04}) reduces to the Kerr metric. Instead, turning off the rotational parameter $a$ followed by a coordinate transformation $r\rightarrow r-Q^2/\mathcal{M}$ transforms the Kerr-Sen solution into the Gibbons-Maeda-Garfinkle-Horowitz-Strominger solution, which describes a static electrically charged black hole in string theory \cite{Gibbons:1987ps,Garfinkle:1990qj}.\\
\indent Now we consider a charged massive scalar particle field $\Psi$ outside of a Kerr-Sen black hole with mass $\mu$ and charge $q$ that obeys the following Klein-Gordon wave equation \cite{Konoplya:2013rxa}:
\be \label{SV01}
\frac{1}{{\sqrt { - g} }}{\partial _\alpha }\left( {{g^{\alpha \beta }}\sqrt { - g} {\partial _\beta }\Psi } \right) - 2iq{A_\alpha }{g^{\alpha \beta }}{\partial _\beta }\Psi  - {q^2}{g^{\alpha \beta }}{A_\alpha }{A_\beta }\Psi  - {\mu ^2}\Psi  = 0.
\ee
To solve the equation above, as usual we can use the ansatz of the scalar field \cite{Brill:1972xj,Teukolsky:1972my,Teukolsky:1973ha,Hartman:2009nz}
\be \label{SV02}
\Psi  = \sum\limits_{l,m} {{\Psi _{lm}}} =\sum\limits_{l,m} {e^{i\left( {m\phi  - \omega t} \right)}}R_{lm}\left( r \right)S_{lm}\left( \theta  \right),
\ee
where $\omega$ is the frequency of the wave field, $l$ is the spheroidal harmonic index, and $m$ is the azimuthal harmonic index. Substituting Eqs. (\ref{KSG09}), (\ref{KSG11}), (\ref{KSG12}), and (\ref{SV02}) into Eq. (\ref{SV01}) gives us two separated equations, specifically the angular part
\be \label{SV03}
\frac{1}{{\sin \theta }}\frac{d}{{d\theta }}\left( {\sin \theta \frac{dS_{lm}\left( \theta  \right)}{{d\theta }}} \right) + \left[ {\lambda_{lm} + {a^2}\left( {{\mu ^2} - {\omega ^2}} \right) - {a^2}{{\cos }^2}\theta \left( {{\mu ^2} - {\omega ^2}} \right) - \frac{{{m^2}}}{{{{\sin }^2}\theta }}} \right]S_{lm}\left( \theta  \right) = 0,
\ee
and the radial part
\be \label{SV04}
\frac{d}{{dr}}\left( {\Delta _{KS}\frac{dR_{lm}\left( r \right)}{{dr}}} \right) + \left[ {\frac{{{G^2}}}{\Delta _{KS}} - {\mu ^2}\left( {{r^2}+ 2br + {a^2}} \right) + 2am\omega  - \lambda_{lm}} \right]R_{lm}\left( r \right) = 0,
\ee
with
\be \label{SV05}
G = \omega \left( {{r^2} + 2br + {a^2}} \right) - qQr - am.
\ee
The coupling constant $\lambda_{lm}$ may be expanded as a
power series \cite{Dolan:2007mj}:
\be \label{SV06}
{\lambda _{lm}} + {a^2}\left( {{\mu ^2} - {\omega ^2}} \right) = l\left( {l + 1} \right) + \sum\limits_{k = 1}^\infty  {{c_k}{a^{2k}}{{\left( {{\mu ^2} - {\omega ^2}} \right)}^k},}
\ee
For later purpose, in the asymptotic regime $l=m\gg1$, the coupling constant (\ref{SV06}) may be expanded as the asymptotic behavior \cite{Hod:2012ib,Hod:2013sna,Hod:2015cqa}:
\be \label{SV07}
{\lambda _{mm}} = {m^2}\left[ {1 + O\left( {{m^{ - 1}}} \right)} \right] - {a^2}\left( {{\mu ^2} - {\omega ^2}} \right).
\ee
\section{Solution in the Near, Far, and Overlap Region}
\label{sec:NFOreg}
\indent Following Refs. \cite{Hartman:2009nz,Hod:2014baa}, we shall outline the new dimensionless variables
\be \label{RE01}
x \equiv \frac{{r - {r_ + }}}{{{r_ + }}},
\ee
\be \label{RE02}
\tau  \equiv \frac{{{r_ + } - {r_ - }}}{{{r_ + }}},
\ee
\be \label{RE03}
k \equiv 2{\omega}{r_ + } - qQ,
\ee
\be \label{RE04}
\delta  \equiv \frac{{\omega  - {\omega _c}}}{{2\pi {T_H}}}.
\ee
Substituting Eqs. (\ref{RE01})--(\ref{RE04}) into Eq. (\ref{SV04}) one finds
\be \label{RE05}
x\left( {x + \tau } \right)\frac{{{d^{2}R(x)}}}{{d{x^2}}} + \left( {2x + \tau } \right)\frac{dR(x)}{{dx}} + VR\left( x \right) = 0,
\ee
where
\be \label{RE06}
V = \frac{{{G^2}}}{{{r_ + }^2x\left( {x + \tau } \right)}} - \lambda  + 2am{\omega} - {\mu ^2}\left[ {{r_ + }^2{{\left( {x + 1} \right)}^2} + 2b{r_ + }\left( {x + 1} \right) + {a^2}} \right],
\ee
\be \label{RE07}
G = r_ + ^2{\omega}{x^2} + {r_ + } x(2b{\omega}+k) +\frac{\delta \tau r_+}{2} .
\ee
We shall consider the following conditions for near-extremal black holes regime with 
\be \label{RE08}
\tau \ll 1,
\ee
and low-frequency scalar-field regime
\be \label{RE09}
\mathcal{M}(\omega_c-\omega) \ll 1.
\ee
For near-extremal Kerr-Sen black holes with conditions (\ref{RE08}) and (\ref{RE09}), there is an overlap region $\tau \ll x \ll 1$; therefore, we can use a matching procedure to determine the complex frequencies which characterize charged massive scalar fields in the background of a charged rotating Kerr-Sen black hole.
\subsection{Near region}
\indent First, we solve the radial Teukolsky equation (\ref{RE05}) within the region
\be \label{RE10}
x\ll 1,
\ee
in which we can approximate the effective radial potential (\ref{RE06}) as
\be \label{RE11}
{V_{near}} \equiv \frac{\left[\left( {2b{\omega} + k} \right)x+\delta\tau/2\right]^2}{{x\left( {x + \tau } \right)}} - \lambda  + 2am{\omega} - {\mu ^2}\left( {{r_ + }^2 + 2b{r_ + } + {a^2}} \right).
\ee
The solution of Eq. (\ref{RE05}) with $V\equiv V_{near}$ is
\bea \nonumber
R\left( x \right) &=& {x^{ - \frac{i}{2}\delta }}{\left( {\frac{x}{\tau } + 1} \right)^{\frac{i}{2}\delta  - i\left( {2b\omega  + k} \right)}}\\\label{RE12}&\quad&\times~_2{F_1}\left( {\frac{1}{2} + i\sigma  - i\left( {2b\omega  + k} \right),\frac{1}{2} - i\sigma  - i\left( {2b\omega  + k} \right);1 - i\delta ; - \frac{x}{\tau }} \right),
\eea
where
\be \label{RE13}
{\sigma ^2} \equiv -\frac{1}{4} - \lambda  + 2am{\omega} + {\left( {2b{\omega} + k} \right)^2} - {\mu ^2}\left( {{r_ + }^2 + 2b{r_ + } + {a^2}} \right).
\ee
It is useful to write Eq. (\ref{RE12}) within the form
\[R\left( x \right) = {x^{ - \frac{i}{2}\delta }}{\left( {\frac{x}{\tau } + 1} \right)^{\frac{i}{2}\delta  - i\left( {2b\omega  + k} \right)}}\left[\frac{{\Gamma \left( {1 - i\delta } \right)\Gamma \left( { - 2i\sigma } \right)}}{{\Gamma \left( {1/2 - i\sigma  - i\left( {2b\omega  + k} \right)} \right)\Gamma \left( {1/2 - i\delta  - i\sigma  + i\left( {2b\omega  + k} \right)} \right)}}\right.\]
\[\left.{ \times _2}{F_1}\left( {\frac{1}{2} + i\sigma  - i\left( {2b\omega  + k} \right),\frac{1}{2} + i\delta  + i\sigma  - i\left( {2b\omega  + k} \right);1 + 2i\sigma ; - \frac{\tau }{x}} \right){\left( {\frac{x}{\tau }} \right)^{ - \frac{1}{2} - i\sigma  + i\left( {2b\omega  + k} \right)}}\right.\]
\[\left. + \frac{{\Gamma \left( {1 - i\delta } \right)\Gamma \left( {2i\sigma } \right)}}{{\Gamma \left( {1/2 + i\sigma  - i\left( {2b\omega  + k} \right)} \right)\Gamma \left( {1/2 - i\delta  + i\sigma  + i\left( {2b\omega  + k} \right)} \right)}}\right.\]
\be \label{RE14}
\left.{ \times _2}{F_1}\left( {\frac{1}{2} - i\sigma  - i\left( {2b\omega  + k} \right),\frac{1}{2} + i\delta  - i\sigma  - i\left( {2b\omega  + k} \right);1 - 2i\sigma ; - \frac{\tau }{x}} \right){\left( {\frac{x}{\tau }} \right)^{ - \frac{1}{2} + i\sigma  + i\left( {2b\omega  + k} \right)}}\right].
\ee
Using the asymptotic limit $_2{F_1}\left( {a,b;c;z} \right) \to 1~{\rm{ for }}~abz/c \to 0$ of the hypergeometric function, one finds the radial solution of the charged massive scalar fields [Eq. (\ref{RE14})] in the overlap region $\tau\ll x \ll 1$ within the form
\be \label{RE15}
R\left( x \right) \to C_1 \times {x^{ - \frac{1}{2} - i\sigma }}+ C_2 \times {x^{ - \frac{1}{2} + i\sigma }},
\ee
where 
\be \label{RE16}
C_1 = \frac{{{\tau ^{1/2 + i\sigma  - i\delta /2}}\Gamma \left( {1 - i\delta } \right)\Gamma \left( { - 2i\sigma } \right)}}{{\Gamma \left( {1/2 - i\sigma  - i\left( {2b\omega  + k} \right)} \right)\Gamma \left( {1/2 - i\delta  - i\sigma  + i\left( {2b\omega  + k} \right)} \right)}},
\ee
\be \label{RE17}
C_2 = \frac{{{\tau ^{1/2 - i\sigma  - i\delta /2}}\Gamma \left( {1 - i\delta } \right)\Gamma \left( {2i\sigma } \right)}}{{\Gamma \left( {1/2 + i\sigma  - i\left( {2b\omega  + k} \right)} \right)\Gamma \left( {1/2 - i\delta  + i\sigma  + i\left( {2b\omega  + k} \right)} \right)}},
\ee
are the normalization constants.
\subsection{Far region}
Next, we solve the radial Teukolsky equation (\ref{RE05}) within the region
\be \label{RE18}
x\gg \tau,
\ee
in which Eq. (\ref{RE05}) can approximately be expressed as
\be \label{RE19}
{x^2}\frac{{{d^{2}R(x)}}}{{d{x^2}}} + 2x\frac{dR(x)}{{dx}} + {{V}_{far}}R\left( x \right) = 0.
\ee
In this far region, the effective radial potential (\ref{RE06}) approaches
\be \label{RE20}
{{V}_{far}} \equiv {\left[ {\left( {2b + r_ + ^{}x} \right){\omega} + k} \right]^2} - \lambda  + 2am{\omega} - {\mu ^2}\left[ {{r_ + }^2{{\left( {x + 1} \right)}^2} + 2b{r_ + }\left( {x + 1} \right) + {a^2}} \right].
\ee
The radial solution of Eq. (\ref{RE19}) is
\bea \nonumber
R\left( x \right) &=& {{C}_3} \times {\left( {2\varepsilon } \right)^{\frac{1}{2} - i\sigma }}{x^{ - \frac{1}{2} - i\sigma }}{e^{ - \varepsilon x}}M\left( {\frac{1}{2} - i\sigma  - \kappa ,1 - 2i\sigma ;2\varepsilon x} \right),\\\label{RE21}
&\quad& +~{{C}_4} \times {\left( {2\varepsilon } \right)^{\frac{1}{2} + i\sigma }}{x^{ - \frac{1}{2} + i\sigma }}{e^{ - \varepsilon x}}M\left( {\frac{1}{2} + i\sigma  - \kappa ,1 + 2i\sigma ;2\varepsilon x} \right),
\eea
where
\be \label{RE22}
\varepsilon  \equiv {r_ + }\sqrt {{\mu ^2} - {\omega}^2},
\ee
\be \label{RE23}
\kappa  \equiv \frac{{\left( {2b{\omega} + k} \right){\omega} - {\mu ^2}\left( {b + {r_ + }} \right)}}{{\sqrt {{\mu ^2} - {\omega}^2} }}
\ee
are the dimensionless variables. Using the asymptotic limit $M\left( {a,b;z} \right) \to 1~{\rm{ for }}~az/b \to 0$ of the Whittaker function \footnote{The Whittaker function $M(a,b;z)$ is the alternative notation for the hypergeometric function~$_1F_1(a,b;z)$}, one finds the radial solution of the charged massive scalar fields [Eq. (\ref{RE19})] in the overlap region 
\be \label{RE24}
\tau\ll x \ll m^{-1},
\ee
within the form
\be \label{RE25}
R(x) \to {{C}_3} \times {\left( {2\varepsilon } \right)^{\frac{1}{2} - i\sigma }}{x^{ - \frac{1}{2} - i\sigma }}+{{C}_4} \times {\left( {2\varepsilon } \right)^{\frac{1}{2} + i\sigma }}{x^{ - \frac{1}{2} + i\sigma }}.
\ee
\subsection{Overlap region}   
For near-extremal Kerr-Sen black holes with the condition (\ref{RE08}), there is an overlap region (\ref{RE24}); therefore, Eqs. (\ref{RE15}) and (\ref{RE25}) can be matched to determine the normalization constants
\be \label{RE26}
{C_3} = \frac{{{{\left( {2\varepsilon } \right)}^{ - 1/2 + i\sigma }}{\tau ^{1/2 + i\sigma  - i\delta /2}}\Gamma \left( {1 - i\delta } \right)\Gamma \left( { - 2i\sigma } \right)}}{{\Gamma \left( {1/2 - i\sigma  - i\left( {2b\omega  + k} \right)} \right)\Gamma \left( {1/2 - i\delta  - i\sigma  + i\left( {2b\omega  + k} \right)} \right)}},
\ee
\be \label{RE27}
{C_4} = \frac{{{{\left( {2\varepsilon } \right)}^{ - 1/2 - i\sigma }}{\tau ^{1/2 - i\sigma  - i\delta /2}}\Gamma \left( {1 - i\delta } \right)\Gamma \left( {2i\sigma } \right)}}{{\Gamma \left( {1/2 + i\sigma  - i\left( {2b\omega  + k} \right)} \right)\Gamma \left( {1/2 - i\delta  + i\sigma  + i\left( {2b\omega  + k} \right)} \right)}}.
\ee
Next, we shall analyze and derive the equation which characterized the complex frequency of charged massive scalar fields in the background of a Kerr-Sen black hole. The radial solution (\ref{RE21}) of the charged massive scalar fields is  characterized by the asymptotic limit $x\to\infty$. The expansion for this asymptotic limit at infinity is given by \cite{Abra}
\[R(x) \to \left[{{C}_3} \times {\left( {2{\varepsilon} } \right)^{\kappa} }{x^{ - 1 + {\kappa} }}{\left( { - 1} \right)^{ - \frac{1}{2} + i{\sigma}  + {\kappa} }}\frac{{\Gamma \left( {1 - 2i{\sigma}  } \right)}}{{\Gamma \left( {\frac{1}{2} - i{\sigma}   + {\kappa} } \right)}}\right.\]
\[\left.+ {{C}_4} \times {\left( {2{\varepsilon} } \right)^{\kappa} }{x^{ - 1 + {\kappa} }}{\left( { - 1} \right)^{ - \frac{1}{2} - i{\sigma}   + {\kappa} }}\frac{{\Gamma \left( {1 + 2i{\sigma}  } \right)}}{{\Gamma \left( {\frac{1}{2} + i{\sigma}   + {\kappa} } \right)}}\right]{e^{ - {\varepsilon} x}}\]
\be \label{RE28}
+ \left[ {{{C}_3} \times {{\left( {2{\varepsilon} } \right)}^{ - {\kappa} }}{x^{ - 1 - {\kappa} }}\frac{{\Gamma \left( {1 - 2i{\sigma}  } \right)}}{{\Gamma \left( {\frac{1}{2} -i {\sigma}   - {\kappa} } \right)}} + {{C}_4} \times {{\left( {2{\varepsilon} } \right)}^{ - {\kappa} }}{x^{ - 1 - {\kappa} }}\frac{{\Gamma \left( {1 + 2i{\sigma}  } \right)}}{{\Gamma \left( {\frac{1}{2} + i{\sigma}   - {\kappa} } \right)}}} \right]{e^{{\varepsilon} x}}.
\ee   
From the boundary condition (\ref{IN03}), we realize that the charged massive scalar fields are characterized by an exponentially decaying radial solution at infinity. This shows that the coefficient of the growing exponent $e^{\varepsilon x}$ must vanish:
\be \label{RE29}
{{{C}_3} \times {{\left( {2{\varepsilon} } \right)}^{ - {\kappa} }}{x^{ - 1 - {\kappa} }}\frac{{\Gamma \left( {1 - 2i{\sigma}  } \right)}}{{\Gamma \left( {\frac{1}{2} -i {\sigma}   - {\kappa} } \right)}} + {{C}_4} \times {{\left( {2{\varepsilon} } \right)}^{ - {\kappa} }}{x^{ - 1 - {\kappa} }}\frac{{\Gamma \left( {1 + 2i{\sigma}  } \right)}}{{\Gamma \left( {\frac{1}{2} + i{\sigma}   - {\kappa} } \right)}}} = 0.
\ee
Plugging Eqs. (\ref{RE26}) and (\ref{RE27}) into Eq. (\ref{RE29}), one finds
\bea \nonumber
&&{\left( {2\varepsilon \tau } \right)^{2i\sigma }}{\left[ {\frac{{\Gamma \left( { - 2i\sigma } \right)}}{{\Gamma \left( {2i\sigma } \right)}}} \right]^2}\\\label{RE30}
&\times&\frac{{\Gamma \left( {\frac{1}{2} + i\sigma  - i\left( {2b\omega  + k} \right)} \right)\Gamma \left( {\frac{1}{2} + i\sigma  - \kappa } \right)\Gamma \left( {\frac{1}{2} - i\delta  + i\sigma  + i\left( {2b\omega  + k} \right)} \right)}}{{\Gamma \left( {\frac{1}{2} - i\sigma  - i\left( {2b\omega  + k} \right)} \right)\Gamma \left( {\frac{1}{2} - i\sigma  - \kappa } \right)\Gamma \left( {\frac{1}{2} - i\delta  - i\sigma  + i\left( {2b\omega  + k} \right)} \right)}} = 1,
\eea
the equation that describes the complex frequencies of the charged massive scalar fields in the background of a near-extremal Kerr-Sen black hole.\\
\indent Equation (\ref{RE30}) can be simplified in the near-extremal Kerr-Sen black holes regime,
\be \label{RE31}
\tau  \ll \frac{{\tilde \omega }}{m}.
\ee
In the near-extremal regime, one can use the approximated gamma function relation that appears in Eq. (\ref{RE30}) as
\be \label{RE32}
\frac{{\Gamma \left( {\frac{1}{2} - i\delta  + i\sigma  + i\left( {2b\omega  + k} \right)} \right)}}{{\Gamma \left( {\frac{1}{2} - i\delta  - i\sigma  + i\left( {2b\omega  + k} \right)} \right)}} = {\left( { - i\delta } \right)^{2i\sigma }}.
\ee
Plugging Eq. (\ref{RE32}) into Eq. (\ref{RE30}), one finds the condition
\be \label{RE33}
{\left( { - 4i\varepsilon \tilde \omega } \right)^{2i\sigma }}{\left[ {\frac{{\Gamma \left( { - 2i\sigma } \right)}}{{\Gamma \left( {2i\sigma } \right)}}} \right]^2}\frac{{\Gamma \left( {\frac{1}{2} + i\sigma  - i\left( {2b\omega  + k} \right)} \right)\Gamma \left( {\frac{1}{2} + i\sigma  - \kappa } \right)}}{{\Gamma \left( {\frac{1}{2} - i\sigma  - i\left( {2b\omega  + k} \right)} \right)\Gamma \left( {\frac{1}{2} - i\sigma  - \kappa } \right)}} = 1,
\ee
where ${\tilde \omega }$ is a dimensionless parameter which quantifies the distance between the frequency of the charged massive scalar field $\omega$ and the critical frequency $\omega_c$ that can be expressed as
\be \label{RE34}
\tilde \omega  \equiv 2\mathcal{M}\left( {\omega  - {\omega _c}} \right),
\ee
and Eq. (\ref{RE33}) is valid in the regime 
\be
m\tau \ll \tilde{\omega} \ll m^{-1}.
\ee
\section{Super-radiant Instability of the Kerr-Sen Black Hole with Charged Massive Scalar Field}
\label{sec:superradiant}
In this section, we shall simplify the condition that describes the complex frequencies of the charged massive scalar fields in the background of a near-extremal Kerr-Sen black hole [Eq. (\ref{RE33})] within the form \footnote{Please note that ``$\times$'' in Eq. (\ref{SI01}) is simply an ordinary multiplication operation.}
\be \label{SI01}
\tilde{\omega}=\mathcal{R} \times \mathcal{I},
\ee
where $\cal R$ and $\cal I$ are, respectively, a real quantity and a complex quantity \footnote{In particular, we use here the identity $1 = {e^{2in\pi}}$. The Eq. (6.1.23) of Ref. \cite{Abra} guarantees that the ratios of the Gamma functions $\Gamma \left( {2i\sigma } \right)/\Gamma \left( { - 2i\sigma } \right) = {e^{i{\phi _1}}}$ and $\Gamma \left( {1/2 - i\sigma  - \kappa } \right)/\Gamma \left( {1/2 + i\sigma  - \kappa } \right) = {e^{i{\phi _2}}}$ in Eq. (\ref{SI02}) are a complex number whose absolute value is 1. This, in turn, implies that the phase $\phi_1$ and $\phi_2$ should be real numbers.},
\be \label{SI02}
\mathcal{R} \equiv \frac{{{e^{ - \pi n/\delta }}}}{{4\varepsilon }}{\left[ {\frac{{\Gamma \left( {2i\sigma } \right)}}{{\Gamma \left( { - 2i\sigma } \right)}}} \right]^{1/i\sigma }}{\left[ {\frac{{\Gamma \left( {\frac{1}{2} - i\sigma  - \kappa } \right)}}{{\Gamma \left( {\frac{1}{2} + i\sigma  - \kappa } \right)}}} \right]^{1/2i\sigma }},
\ee
\be \label{SI03}
\mathcal{I} \equiv i{\left[ {\frac{{\Gamma \left( {\frac{1}{2} - i\sigma  - i\left( {2b\omega  + k} \right)} \right)}}{{\Gamma \left( {\frac{1}{2} + i\sigma  - i\left( {2b\omega  + k} \right)} \right)}}} \right]^{1/2i\sigma }}.
\ee
From Eqs. (\ref{SI01})--(\ref{SI03}), one finds relations
\be \label{SI04}
{{\tilde \omega }_R} = \mathcal{R} \times \mathcal{I}_R,
\ee
\be \label{SI05}
{{\tilde \omega }_I} = \mathcal{R} \times \mathcal{I}_I,
\ee
which can be simplified in a dimensionless ratio \footnote{Please note that Eq. (\ref{SI06}) follows from Eq. (\ref{SI05}) with the help of Eqs. (\ref{RE34}) and (\ref{SI02}) which implies that $\cal R$ is a real quantity.},
\be \label{SI06}
\frac{{{\omega _I}}}{{{\omega _R} - {\omega _c}}} = \frac{{{\mathcal{I}_I}}}{{{\mathcal{I}_R}}},
\ee
for the frequencies of the charged massive scalar fields in a near-extremal Kerr-Sen black hole.

\section{The Eikonal Large-Mass Regime}
\label{sec:eikonal}
In this section, we study the eikonal large-mass regime of the charged massive scalar fields in the background of a Kerr-Sen black hole. In the asymptotic large-mass regime,
\be \label{ELM01}
\mathcal{M}\mu\gg 1,
\ee
the ratio of the gamma function in Eq. (\ref{SI03}) can be approximated by \cite{Abra}
\bea \nonumber
\frac{{\Gamma \left( {\frac{1}{2} - i\sigma  - i\left( {2b\omega  + k} \right)} \right)}}{{\Gamma \left( {\frac{1}{2} + i\sigma  - i\left( {2b\omega  + k} \right)} \right)}} &=& {e^{\sigma \left( {2i - \pi } \right)}}{\left( {2b\omega  + k + \sigma } \right)^{ - i\left( {2b\omega  + k + \sigma } \right)}}{\left( {2b\omega  + k - \sigma } \right)^{i\left( {2b\omega  + k - \sigma } \right)}}\\\label{ELM02}
&\quad& \times \left[ {1 + {e^{ - 2\pi \left( {2b\omega  + k - \sigma } \right)}}} \right]\left[ {1 + O\left( {{m^{ - 1}}} \right)} \right].
\eea
Substituting Eq. (\ref{ELM02}) into Eq. (\ref{SI03}), one finds \footnote{Here, we use the relation $ie^{-\pi/2i}=-1$. In the asymptotic $2b\omega+k-\sigma\gg 1$ regime, we use the approximated relation $\left[ {1 + {e^{ - 2\pi \left( {2b\omega  + k - \sigma } \right)}}} \right]=1 - i{e^{ - 2\pi \left( {2b\omega  + k - \sigma } \right)}}/2\sigma  + O\left[ {{{\left( {{e^{ - 2\pi \left( {2b\omega  + k - \sigma } \right)}}/\sigma } \right)}^2}} \right]$.}
\be \label{ELM03}
{\mathcal{I}_R} =  - e{\left( {2b\omega  + k + \sigma } \right)^{ - \left( {2b\omega  + k + \sigma } \right)/2\sigma }}{\left( {2b\omega  + k - \sigma } \right)^{\left( {2b\omega  + k - \sigma } \right)/2\sigma }},
\ee
\be \label{ELM04}
{\mathcal{I}_I} = {\mathcal{I}_R} \times \frac{{ - {e^{ - 2\pi \left( {2b\omega  + k - \sigma } \right)}}}}{{2\sigma }}.
\ee
Substituting Eqs. (\ref{ELM03}) and (\ref{ELM04}) into Eq. (\ref{SI06}), one finds a quite simple dimensionless relation,
\be \label{ELM05}
\frac{{{\omega _I}}}{{{\omega _R} - {\omega _c}}} = \frac{{ - {e^{ - 2\pi \left( {2b\omega  + k - \sigma } \right)}}}}{{2\sigma }},
\ee
for the frequencies of charged massive scalar field in the background of a Kerr-Sen black hole in the eikonal large-mass regime. It turns out that Eq. (\ref{ELM05}) for Kerr-Sen spacetime resembles the equation for the Kerr-Newman spacetime as found by Hod in \cite{Hod:2016bas}. However there are several distinguishable parameters such as $b$ and $\sigma$.\\ 
\indent Following \cite{Hod:2016bas}, next, we compare the large-mass result [Eq. (\ref{ELM05})] with the famous Wentzel-Kramers-Brillouin (WKB) result of Zouros and Eardly for the neutral scalar fields coupled to a rotating Kerr black hole in the (\ref{ELM01}) regime \cite{Zouros:1979iw}:
\be \label{ELM06}
\mathcal{M}{\omega _I} \sim {e^{ - 2\pi \left( {2 - \sqrt 2 } \right)\mathcal{M}\mu }},
\ee
under these conditions
\be \label{ELM07}
a \simeq \mathcal{M},
\ee
\be \label{ELM08}
l = m \gg 1,
\ee
\be \label{ELM09}
\omega  \simeq \mu  \simeq \frac{m}{{2\mathcal{M}}} \gg 1,
\ee
for near-extremal Kerr black holes. Substituting Eqs. (\ref{ELM07})--(\ref{ELM09}) into Eqs. (\ref{SV07}), (\ref{RE03}), and (\ref{RE13}), one finds
\be \label{ELM10}
k = m,
\ee
\be \label{ELM11}
\sigma  = \frac{{m\sqrt 2 }}{2} + O\left( 1 \right).
\ee
By setting $b=0$, the line element (\ref{KSG04}) reduces to the Kerr metric, and plugging back Eqs. (\ref{ELM10}) and (\ref{ELM11}) to our result [Eq. (\ref{ELM05})], one finds
\be
\frac{{{\omega _I}}}{{{\omega _c} - {\omega _R}}} = \frac{{{e^{ - 2\pi \left( {2 - \sqrt 2 } \right)\mathcal{M}\mu }}}}{{2\sqrt 2 \mathcal{M}\mu }},
\ee
which is consistent with the WKB result of Eq. (\ref{ELM06}) for the neutral scalar fields coupled to a rotating Kerr black hole in the eikonal large-mass regime.
\section{Summary and Discussion}
In this paper, we have studied analytically the super-radiant instability properties of the charged massive scalar field in a Kerr-Sen black hole background by analyzing the complex resonance parameter which characterizes the charged massive scalar field in the Kerr-Sen black hole spacetime. In the eikonal large-mass $\mathcal{M}\gg \mu$ regime, it is found that the dimensionless ratio for the frequencies of the charged massive scalar fields in the near-extremal Kerr-Sen black hole spacetime is also consistent with the result of Zouros and Eardly \cite{Zouros:1979iw} for the case of neutral scalar fields in the background of a near-extremal Kerr black hole. In showing this consistency, we adopt the method by Hod in Ref. \cite{Hod:2016bas}, where the case of the Kerr-Newman black hole spacetime is studied. As we know, both Kerr-Newman and Kerr-Sen black holes have quite similar physical properties, and because both are rotating and charged black hole solutions, it is natural to expect that the result resembles the Kerr-Newman one.\\
\indent Finally, we would like to note that it is interesting to analyze the super-radiant instability growth rate [Eq. (\ref{ELM05})] in the dimensionless charge-to-mass ratio, which characterized the explosive charged massive scalar fields. For the composed Kerr-Newman black-hole-charged massive-scalar-field system, it was shown by Hod in Ref. \cite{Hod:2016bas} that, in the eikonal large-mass $\mathcal{M}\gg \mu$ regime, the super-radiant instability growth rates of the explosive charged massive scalar fields are characterized by a nontrivial dependence on the dimensionless charge-to-mass ratio. Intuitively, we predict that the same outcome can be found for the case of the composed Kerr-Sen black-hole-charged massive-scalar-field system.
\begin{acknowledgments}
I would like to thank Haryanto M. Siahaan for suggesting this study and for fruitful discussions. I would like to thank Professor Shahar Hod, Professor Carlos Herdeiro, and the anonymous referees for reading this manuscript and for their useful comments and constructive suggestions. I also thank the Department of Physics, Parahyangan Catholic University, for the support and encouragement.
\end{acknowledgments}
\bibliography{apssamp}
\end{document}